\documentclass{webofc}
\usepackage[varg]{txfonts}   
\usepackage{epstopdf}
\begin{document}
\title{Antiproton--nucleus quasi-bound states within \\ the 2009 version of the Paris $\bar{N}N$ potential}
%
%

\author{\firstname{Jaroslava} \lastname{Hrtánková}\inst{1}\fnsep\thanks{\email{hrtankova@ujf.cas.cz}} \and
        \firstname{Ji\v{r}í} \lastname{Mare\v{s}}\inst{1}}
       

\institute{Nuclear Physics Institute, 250 68 \v{R}e\v{z}, Czech Republic
          }

\abstract{%
 We studied the ${\bar p}$ interactions with the nuclear medium within the 2009 version of the Paris ${\bar N}N$ potential model. We constructed the $\bar{p}$--nucleus optical potential using the Paris $S$- and $P$-wave ${\bar p}N$ scattering amplitudes and treated their strong energy and density dependence self-consistently. We considered a phenomenological $P$-wave term as well. We calculated $\bar{p}$ binding energies and widths of the $\bar{p}$ bound in various nuclei. The $P$-wave potential has very small effect on the calculated ${\bar p}$ binding energies, however, 
it reduces the corresponding widths noticeably. Moreover, the $S$-wave potential based on the Paris amplitudes supplemented 
by a phenomenological $P$-wave term yields the ${\bar p}$ binding energies and widths in very good agreement with those 
obtained within the RMF model consistent with ${\bar p}$-atom data. 
}
\maketitle
\section{Introduction}
\label{intro}

The antiproton--nucleus interaction below threshold have been so far studied within phenomenological RMF approaches~\cite{Mishustin, hmNPA16}. The G-parity motivated $\bar{p}$ coupling constants were used to construct the $\bar{p}$--nucleus potential. The absorption of $\bar{p}$ was accounted for in terms of a purely phenomenological optical potential. The $\bar{p}$ optical potential was confronted with $\bar{p}$ atom data. It was found that the $\bar{p}$ coupling constant have to be properly scaled in order to be consistent with the data. Consequently, the $\bar{p}$ potential was applied in the calculations of $\bar{p}$ quasi-bound states in various nuclei~\cite{hmNPA16}. 

However, it is desirable to study the $\bar{p}$ interactions with the nuclear medium within other theoretical approaches, such as microscopic models of $\bar{N}N$ interaction based on meson-exchange models~\cite{paris09, Bonn, ZT} or chiral $\bar{N}N$ interaction models~\cite{KHM14, H17}. Comparison between these ${\bar N}N$ interaction models could bring valuable information about in-medium $\bar{p}$ interactions in the direct confrontation with the data from $\bar p$ atoms and $\bar{p}$ scattering off nuclei, as well as predictions for $\bar{p}$-nuclear quasi-bound states. 

Recently, the 2009 version of the Paris ${\bar N}N$ potential~\cite{paris09} was confronted by Friedman \emph{et al.} with the ${\bar p}$-atom data and antinucleon interactions with nuclei up to 400~MeV/c, including elastic scattering and annihilation 
cross sections~\cite{friedmanNPA15}. The analysis revealed the necessity to include the $P$-wave interaction in order to describe the $\bar{p}$ atom data. The Paris $S$-wave potential supplemented by a phenomenological $P$-wave term was found to fit the data on low-density, near-threshold ${\bar p}$-nucleus interaction.  This fact stimulated us to apply it in the present calculations of $\bar{p}$-nuclear quasi-bound states and explore the effect of the $P$-wave interaction on $\bar{p}$ binding energies and widths of $\bar{p}$-nuclear states.

In Section 2, we briefly introduce the model applied in our calculations. Section 3 presents few representative results together with the discussion of the main findings of our study.

\section{Methodology}
\label{sec-1}

The binding energies $B_{\bar{p}}$ and widths $\Gamma_{\bar{p}}$ of $\bar{p}$ quasi-bound states in a nucleus are obtained 
by solving self-consistently the Dirac equation with the optical potential
\begin{equation} \label{DiracEqVopt}
[-i \alpha \cdot \nabla +\beta m_{\bar{p}}+V_{\text{opt}}(r)]\psi_{\bar{p}}=\epsilon_{\bar{p}} \psi_{\bar{p}},
\end{equation}
where $m_{\bar{p}}$ is the mass of the antiproton and $\epsilon_{\bar{p}}=-B_{\bar{p}} - {\rm i}\Gamma_{\bar{p}}/{2}$ ($B_{\bar{p}}>0$). The $S$-wave $\bar{p}$--nucleus optical potential $V_{\rm opt}$ enters the Dirac equation as the time component of a 4-vector and is constructed in a `$t\rho$' form as follows: 
\begin{equation} \label{SoptPot}
 2E_{\bar{p}}V_{\text{opt}}(r)=-4\pi \left(F_0\frac{1}{2}\rho_p(r) 
+ F_1\left(\frac{1}{2}\rho_p(r)+\rho_n(r) \right)\right)~.
\end{equation}
Here, $E_{\bar{p}}=m_{\bar{p}}-B_{\bar{p}}$, $F_0$ and $F_1$ are isospin 0 and 1 in-medium amplitudes, and $\rho_p(r)$ [$\rho_n(r)$] is the proton (neutron) density distribution calculated 
within the RMF NL-SH model \cite{nlsh}. 
The in-medium amplitudes $F_0$ and $F_1$ entering Eq.~\eqref{SoptPot} account for Pauli correlations in the nuclear medium. They are constructed from the free-space $\bar{p}N$ amplitudes 
using the multiple scattering approach of Wass \emph{et al.} \cite{wrw} (WRW) 
\begin{equation}\label{wrwAmp}
F_{1}=\frac{\frac{\sqrt{s}}{m_N}f^{\rm S}_{\bar{p}n}(\sqrt{s})}{1\!+\!\frac{1}{4}\xi_k \frac{\sqrt{s}}{m_N} f^{\rm S}_{\bar{p}n}
(\sqrt{s}) \rho (r)}~,\;\; F_{0}=\frac{\frac{\sqrt{s}}{m_N}[2f^{\rm S}_{\bar{p}p}(\sqrt{s})\! -\! f^{\rm S}_{\bar{p}n}
(\sqrt{s})]}{1\!+\!\frac{1}{4}\xi_k \frac{\sqrt{s}}{m_N}[2f^{\rm S}_{\bar{p}p}(\sqrt{s})\!  
-\! f^{\rm S}_{\bar{p}n}( \sqrt{s})] \rho (r)}~. 
\end{equation} 
Here, $f^{\rm S}_{\bar{p}n}$ ($f^{S}_{\bar{p}p}$) denotes the free-space c.m. $\bar{p}n$ (${\bar{p}p}$) $S$-wave scattering amplitude derived from the Paris $\bar{N}N$ potential as a function of Mandelstam variable $\sqrt{s}$, $m_N$ represents the mass of the nucleon and $\rho(r) = \rho_p(r)+\rho_n(r)$. The factor $\sqrt{s}/m_N$ transforms the amplitudes from the two-body frame to the $\bar{p}$--nucleus frame. The Pauli correlation factor $\xi_k$ is defined as follows
\begin{equation} \label{xi}
 \xi_k=\frac{9\pi}{k_{\rm F}^2}\,\left(4\! \int_0^{\infty} \frac{dr}{r} \exp(ikr)j_1^2(k_{\rm F} r)\right)\; ,
\end{equation}
where $j_1(k_{\rm F} r)$ is the spherical Bessel function, $k_{\rm F}$ is the Fermi momentum and $k=~\sqrt{(\epsilon_{\bar{p}}+m_{\bar{p}})^2-m_{\bar{p}}^2}$ is the antiproton momentum.
The integral in Eq.\eqref{xi} can be solved analytically. The resulting expression is of the form
\begin{equation}
 \xi_k=\frac{9\pi}{k_{\rm F}^2} \left[1- \frac{q^2}{6}+\frac{q^2}{4}\left(2+ \frac{q^2}{6}\right)
\ln\left(1+\frac{4}{q^2}\right)- \frac{4}{3}q \left( \frac{\pi}{2}- \arctan \left( \frac{q}{2} \right) \right) \right]~,
\end{equation}
where $q=-ik/k_{\rm F}$.

\begin{figure}[t!]
\begin{center} 
\includegraphics[width=0.4\textwidth]{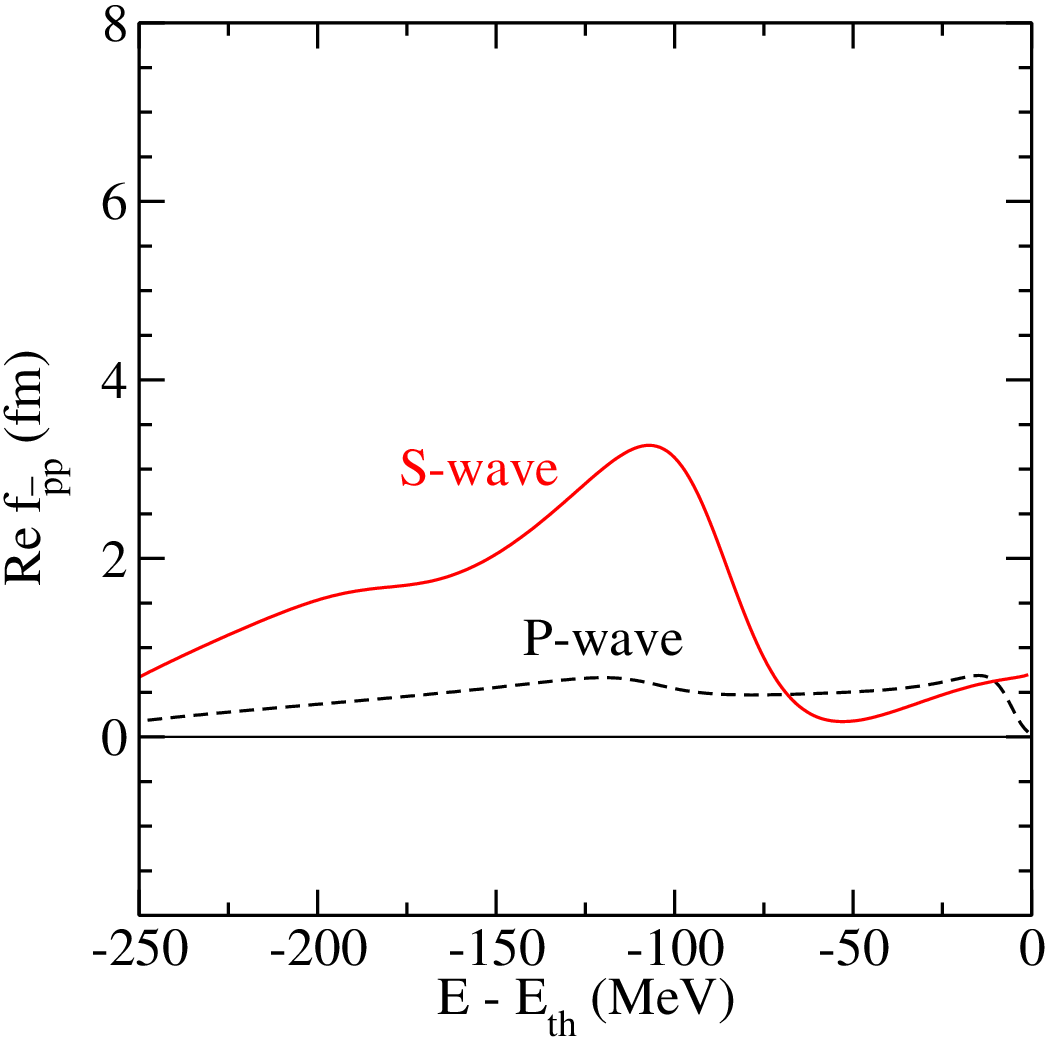} \hspace{5pt}
\includegraphics[width=0.4\textwidth]{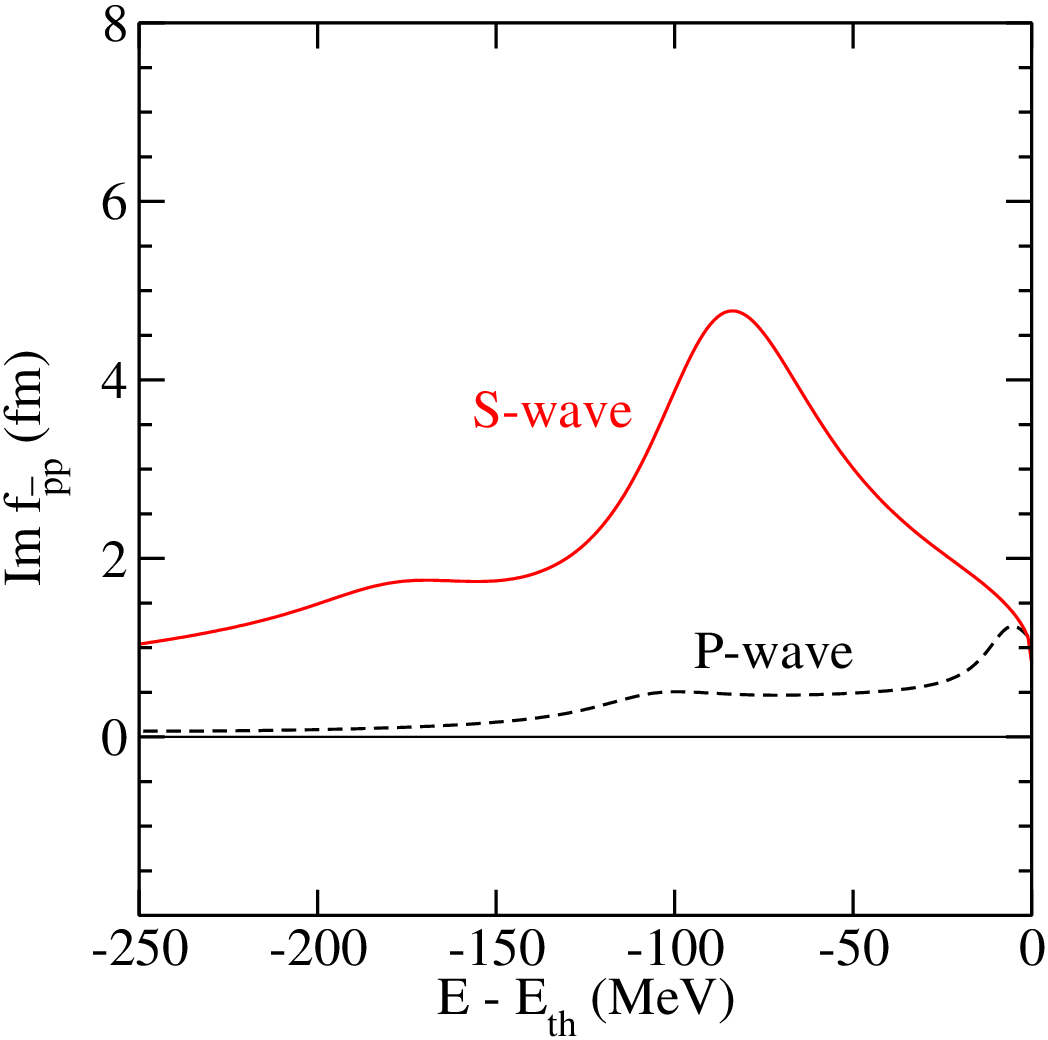} \\[5pt]
\includegraphics[width=0.4\textwidth]{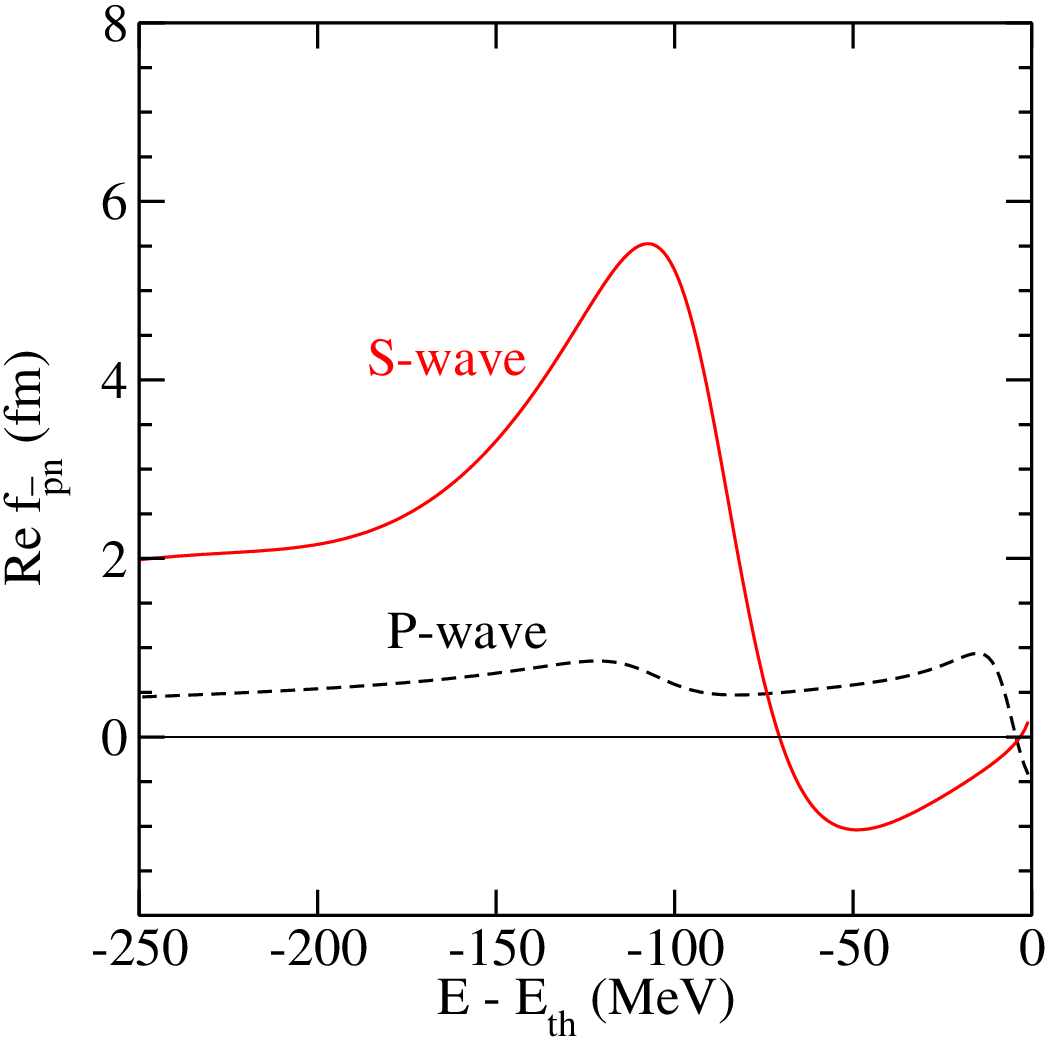} \hspace{5pt}
\includegraphics[width=0.4\textwidth]{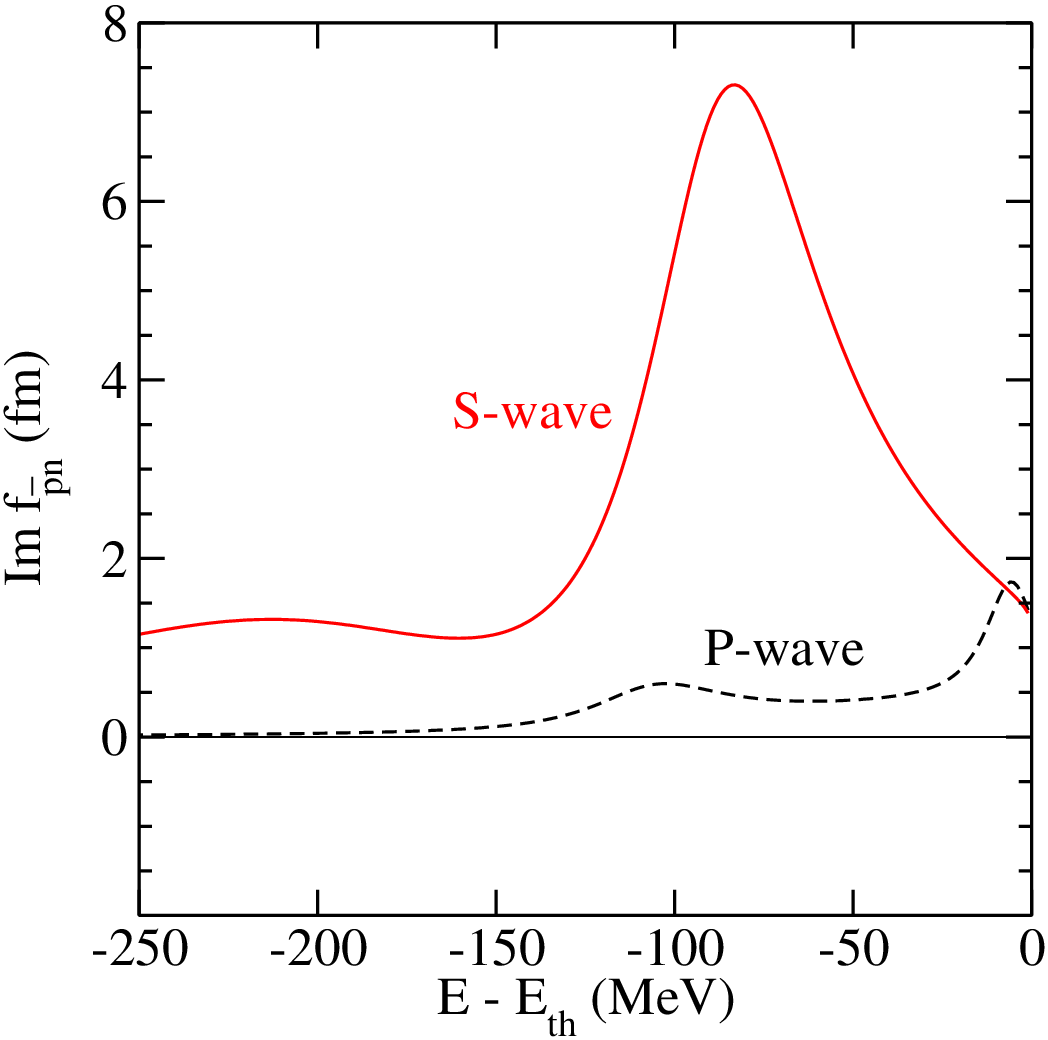}
\caption{\label{fig-1}Energy dependence of real (left) and imaginary (right) parts of the Paris 09 $\bar{p}p$ (top) and $\bar{p}n$ (bottom) two-body c.m. scattering 
amplitudes used in the present calculations: in-medium (Pauli blocked) $S$-wave amplitudes at $\rho_0=0.17$~fm$^{-3}$ and free-space $P$-wave amplitudes.}
\end{center}
\end{figure}

The analysis of $\bar{p}$ atom data~\cite{friedmanNPA15} revealed that it is necessary to supplement the Paris $S$-wave potential by the $P$-wave interaction to make the real $\bar{p}$ potential attractive in the relevant low-density region of a nucleus. To incorporate the $P$-wave interaction in our model we supplement the r.h.s. of the $S$-wave optical potential in Eq.~\eqref{SoptPot} [$2E_{\bar{p}}V_{\rm opt}^S=q(r)$] by a gradient term~\cite{friedmanNPA15}:
\begin{equation}\label{SPpot}
 2E_{\bar{p}}V_{\text{opt}}(r)=q(r)+3 \nabla \cdot \alpha(r) \nabla\; .
\end{equation}
The factor $2l+1 = 3$ in the $P$-wave part is introduced to match the normalization of the Paris $\bar{N}N$ 
scattering amplitudes and 
\begin{equation}
 \alpha(r) = 4 \pi \frac{m_N}{\sqrt{s}}\left(f^{P}_{\bar{p}p}( \sqrt{s}) \rho_p(r) + 
f^{P}_{\bar{p}n}( \sqrt{s}) \rho_n(r) \right)~.
\end{equation}
Here, $f^P_{\bar{p}p}( \sqrt{s})$ and $f^P_{\bar{p}n}( \sqrt{s})$ represent the Paris $P$-wave 
$\bar{p}p$ and $\bar{p}n$ free-space c.m. scattering amplitudes, respectively. We do not consider any medium modifications of the $P$-wave amplitudes since we assume that the $P$-wave potential should contribute mainly near the surface of the nucleus due to its gradient form. 

The analysis of Ref.~\cite{friedmanNPA15} also revealed that the optical potential constructed from the Paris $S$- and $P$-wave amplitudes fails to reproduce the $\bar{p}$ atom data and that it is mainly due to the $P$-wave amplitude --- its real and imaginary parts had to be scaled by different factors to get reasonable fit. On the contrary, the optical potential based on the Paris $S$-wave potential supplemented by a purely phenomenological $P$-wave term with $f^P_{\bar{p}N}=2.9+i1.8$~fm$^3$ fits the data well. In our calculations, we adopt both $P$-wave amplitudes, Paris as well as phenomenological, in order to study their effect on the binding energies and widths of $\bar{p}$-nuclear states.

The Paris amplitudes used in our calculations are shown in Fig.~\ref{fig-1}. There are $\bar{p}p$ (top) and $\bar{p}n$ (bottom) medium modified $S$-wave amplitudes \eqref{wrwAmp} at saturation density $\rho_0=0.17$~fm$^{-3}$ and free-space $P$-wave scattering amplitudes plotted as a function of the energy shift $\delta \sqrt{s}=E-E_{\rm th}$ with $E_{\rm th}=m_{\bar p}+m_N$. The $S$-wave amplitudes vary considerably with energy below threshold. The real in-medium $\bar{p}p$ amplitude is attractive in the entire energy region below threshold. The real part of the in-medium $\bar{p}n$ amplitude is attractive for $\delta \sqrt{s}\leq-70$~MeV with a small repulsive dip near threshold. The imaginary parts of the $S$-wave amplitudes are comparable or even larger than the corresponding real parts. The energy dependence of the free-space $P$-wave amplitudes is less pronounced than in the $S$-wave case. Moreover, the $P$-wave amplitudes are considerably smaller than the in-medium $S$-wave amplitudes in the region relevant to $\bar{p}$-nuclear states calculations.

Strong energy dependence of the $\bar{p}N$ amplitudes presented in Fig.~\ref{fig-1} requires a proper self-consistent scheme for evaluating the $\bar{p}$ optical potential. The energy argument $\sqrt{s}$ of the amplitudes is expressed in the $\bar{p}$--nucleus frame where the contributions from antiproton and nucleon kinetic energies are not negligible~\cite{cfggmPLB}
\begin{equation} \label{Eq.:J}
 \sqrt{s}= E_{\rm th} \left(\!1-\frac{2(B_{\bar{p}} + B_{Nav})}{E_{\rm th}} + \frac{(B_{\bar{p}}+ B_{Nav})^2}{E_{\rm th}^2} - \frac{T_{\bar{p}}}{E_{\rm th}} - \frac{T_{Nav}}{E_{\rm th}} \!\right)^{1/2}~.
\end{equation}
Here, $B_{Nav}=8.5$~MeV and $T_{Nav}$ are the average binding and kinetic energy per nucleon, respectively,
and $T_{\bar{p}}$ represents the $\bar{p}$ kinetic energy. The kinetic energies are evaluated as corresponding expectation values of the kinetic energy operator $\hat{T}=-\frac{\hbar^2}{2 m} \triangle$. Since the $B_{\bar{p}}$ appears as an argument in the $\sqrt{s}$, which in turn serves as an argument for $V_{\rm opt}$, $\sqrt{s}$ has to be determined self-consistently. Namely, its value obtained by solving Eq.~\eqref{Eq.:J} should agree with the value of $\sqrt{s}$ which serves as input in Eq.~\eqref{wrwAmp} and thus Eq.~\eqref{DiracEqVopt}, as well.

\section{Results}
\label{sec-2}
We performed self-consistent calculations of $\bar{p}$-nuclear quasi-bound states in selected nuclei within the model presented in the previous section. We explored the energy and density dependence of the $S$-wave $\bar{p}$--nucleus potential as well as the role of the $\bar{p}N$ $P$-wave interaction, and compared the predictions for $\bar{p}$ binding energies and widths with the phenomenological RMF approach~\cite{hmNPA16}.

\begin{figure}[t!]
\begin{center}
\includegraphics[width=0.8\textwidth]{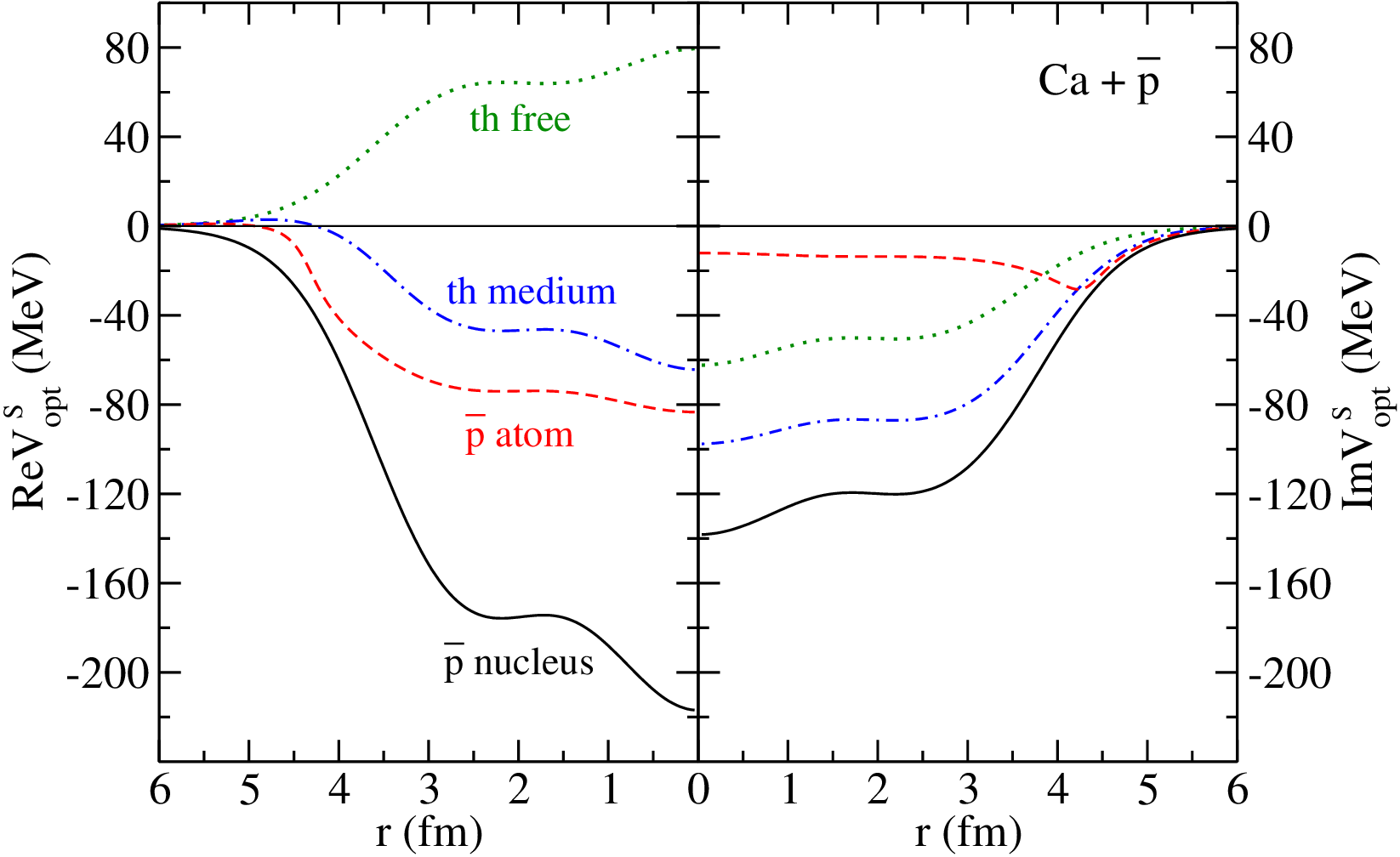}
\caption{\label{fig-2}The potential felt by $\bar{p}$ at threshold (`th medium'), in the $\bar{p}$ atom and $\bar{p}$ nucleus, 
calculated for $^{40}$Ca+$\bar{p}$ with in-medium Paris $S$-wave amplitudes and static RMF densities. 
The ${\bar p}$ potential calculated using free-space amplitudes at threshold is shown for comparison (`th free').}
\end{center}
\end{figure}

The $\bar{p}N$ amplitudes are strongly energy and density dependent, as was shown in Fig.~\ref{fig-1}. Consequently, 
the depth and shape of the $\bar{p}$--nucleus potential depend greatly on the energies and densities 
pertinent to the processes under consideration. It is demonstrated in Fig.~\ref{fig-2} where we 
present the $\bar{p}$ potential in $^{40}$Ca calculated for different energies and densities: i) using the Paris free-space $S$-wave amplitudes at threshold (denoted by `th free'), ii) 
using in-medium Paris $S$-wave amplitudes at threshold (denoted by `th medium'), iii) using in-medium Paris $S$-wave amplitudes at energies relevant to ${\bar p}$ atoms (constructed following Ref.~\cite{friedmanNPA15}), and iv) using in-medium Paris $S$-wave amplitudes at energies relevant to $\bar{p}$ nuclei [$\sqrt{s}$ of Eq.~\eqref{Eq.:J}]. The $\bar{p}$ potential constructed using the free-space amplitudes has a repulsive real 
part and fairly absorptive imaginary part. When the medium modifications of the amplitudes are taken into account, the $\bar{p}$ potential becomes attractive and more absorptive. At the energies relevant to ${\bar p}$ atoms, the $\bar{p}$ potential is more attractive and weakly absorptive. Finally, at the energies relevant to $\bar{p}$ nuclei, the $\bar{p}$ potential is strongly attractive, however, 
also strongly absorptive. The figure clearly shows that proper self-consistent evaluation of the energy $\sqrt{s}$ is essential. 

\begin{figure}[t!]
\begin{center}
\includegraphics[width=0.48\textwidth]{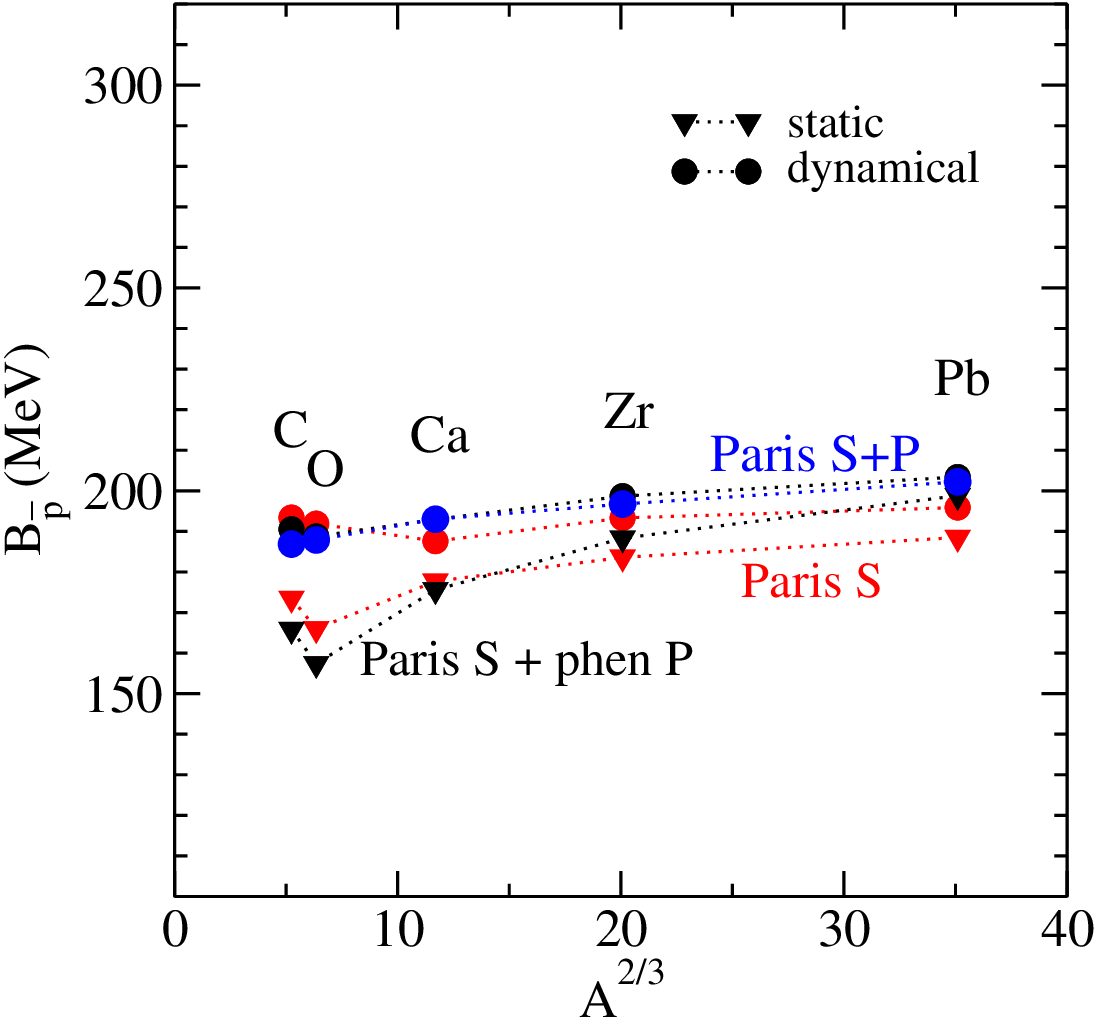} \hspace{5pt}
\includegraphics[width=0.48\textwidth]{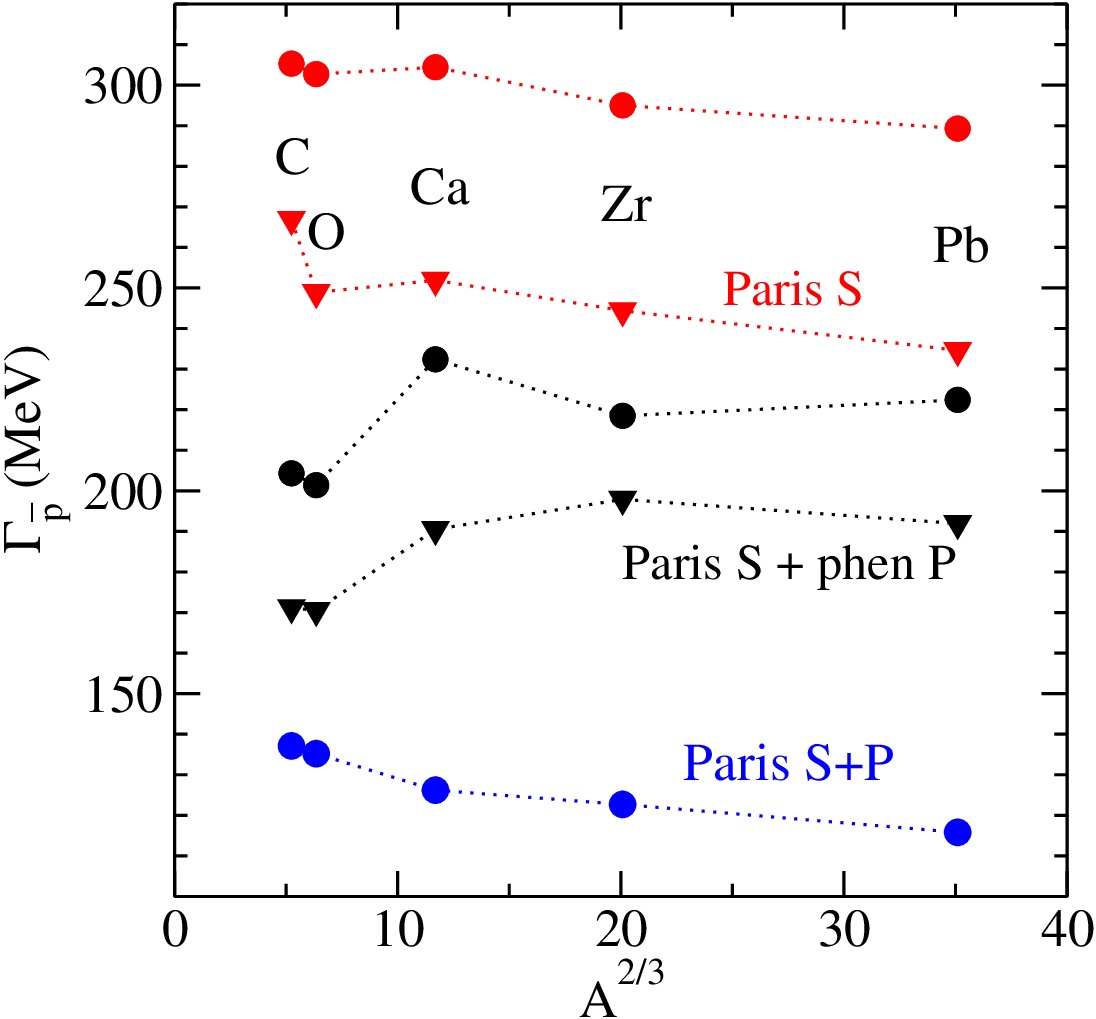}
\caption{\label{fig-5}$1s$ $\bar{p}$ binding energies (left panel) and widths (right panel) in various nuclei, calculated
statically (triangles) and dynamically (circles) using $S$-wave Paris potential (red)  
and including phenomenological $P$-wave potential (black). The $\bar{p}$ binding energies and widths calculated dynamically using the Paris $S+P$-wave potential (blue circles) are shown for comparison.}
\end{center}
\end{figure}

Next, we performed static and dynamical calculations of $\bar{p}$ binding energies and widths using the Paris $\bar{N}N$ potential. In the static calculations, the core nucleus is not affected by the presence of extra $\bar{B}$. In the dynamical calculations, the polarization of the nuclear core due to $\bar{B}$, i.e., changes in the nucleon binding energies and densities, is taken into account. The response of the nuclear core to the extra antiproton is not instant --- it could possibly last longer than the lifetime of $\bar{p}$ inside a nucleus \cite{larionovPRC08, lmsgPRC10}. As a result, the antiproton could annihilate before the nuclear core is fully polarized. Our static and dynamical calculations of $\bar{p}$ binding energies and widths may be thus considered as two limiting scenarios.

In Fig.~\ref{fig-5}, we present $1s$ $\bar{p}$ binding energies (left) and widths (right) as a function of 
mass number $A$, calculated statically (triangles) and dynamically (circles) with the Paris $S$-wave and 
Paris $S$-wave + phen. $P$-wave potentials. We present the $\bar{p}$ binding energies and widths calculated dynamically using the Paris $S+P$-wave potential for comparison as well.  

In dynamical and static calculations alike, the $P$-wave interaction does not affect much the $\bar{p}$ binding 
energies --- they are comparable with the binding energies evaluated using only the $S$-wave potential. 
On the other hand, the $\bar{p}$ widths are reduced significantly when the phenomenological $P$-wave term is 
included in the ${\bar p}$ optical potential. The effect is even more pronounced for the Paris $P$-wave interaction. 

The $\bar{p}$ widths calculated dynamically are noticeably larger than the widths calculated statically. It is caused by the increase of the central nuclear density, which exceeds the decrease of the 
${\bar p}N$ amplitudes due to the larger energy shift with respect to threshold ($\delta \sqrt{s} \sim -255$~MeV in the dynamical case vs. $\delta \sqrt{s} \sim -200$~MeV in the static case). On the other hand, the $\bar{p}$ binding energies increase only moderately and get closer to each other when the dynamical effects are taken into account.
The $\bar{p}$ widths exhibit much large dispersion then the $\bar{p}$ binding energies for the different potentials.

\begin{figure}[b!]
\begin{center}
\includegraphics[width=0.94\textwidth]{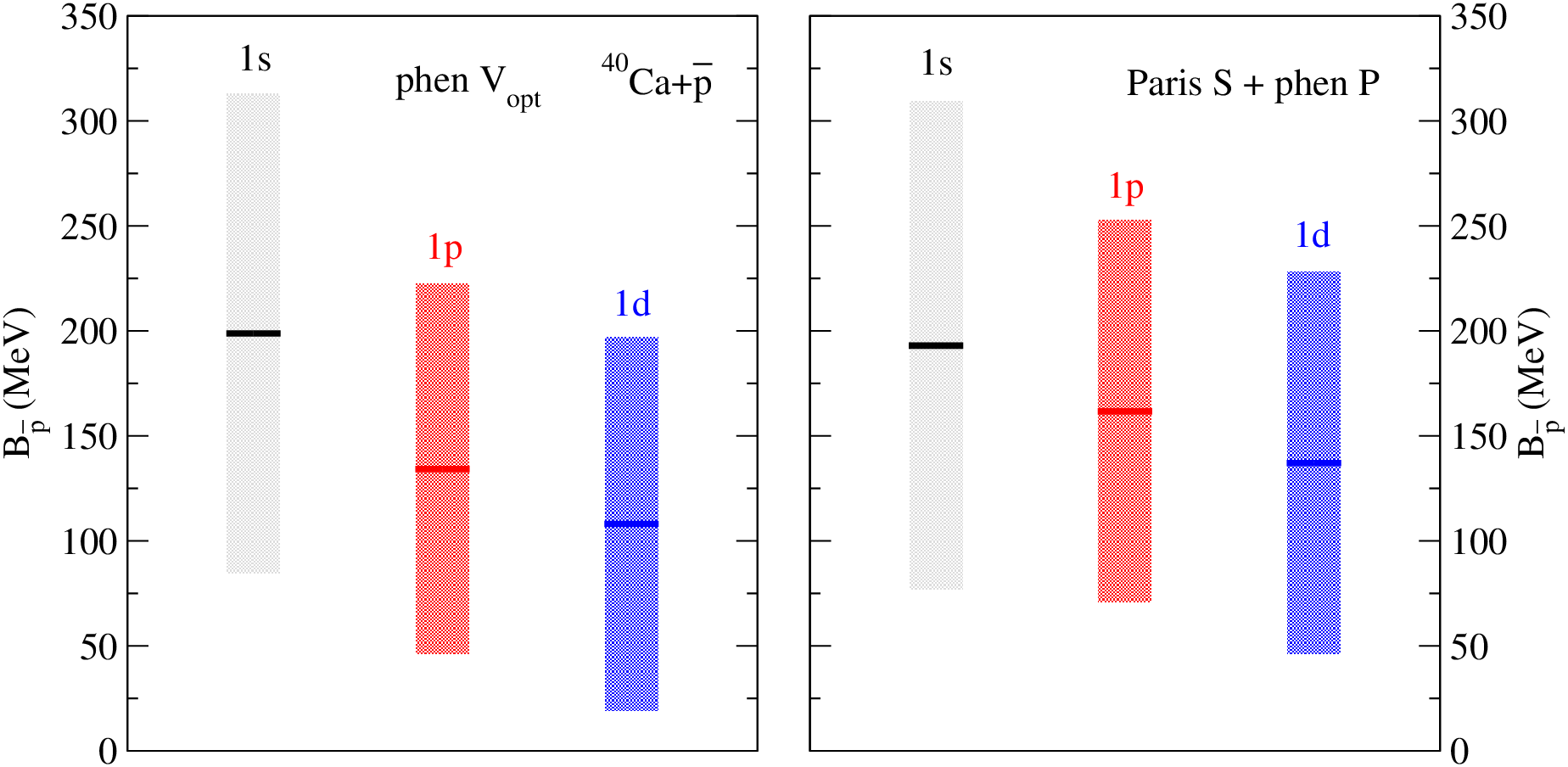}
\caption{\label{fig-4}$1s$, $1p$ and $1d$ binding energies (lines) and widths (boxes) of $\bar{p}$ in $^{40}$Ca calculated 
dynamically within the phenomenological RMF $\bar{p}$ optical potential and Paris $S$-wave + phen. $P$-wave potential.}
\end{center}
\end{figure}
We explored the $\bar{p}$ excited states in selected nuclei as well and compared the results with those obtained within the RMF approach~\cite{hmNPA16}. Fig.~\ref{fig-4} shows ${\bar p}$ spectra in $^{40}$Ca calculated using the Paris $S$-wave + phen. $P$-wave potential and phenomenological RMF approach. The Paris $S$-wave + phen. $P$-wave potential yields the $1p$ and $1d$ binding energies slightly larger and thus the $s\text{-}p$ and $s\text{-}d$ level spacing smaller than the RMF approach. It is an effect of a broader $\bar{p}$ potential well 
generated by the Paris $S$-wave + phen. $P$-wave potential. Nevertheless, both approaches yield comparable $\bar{p}$ widths as well as energies and the overall agreement is surprisingly good. 

It is to be noted that there is no spin-orbit splitting of the $p$ and $d$ levels presented in Fig.~\ref{fig-4} since the $V_{\rm opt}$ is a central potential constructed from angular momentum-averaged scattering amplitudes. 
In the RMF approach, the $\bar{p}$ binding energies in $1p$ and $1d$ spin doublets are nearly degenerate, the difference in $\bar{p}$ energies (as well as $\bar{p}$ widths) is up to~$\sim 1$~MeV. This is in agreement with spin symmetry in
antinucleon spectra within the RMF approach~\cite{ginocchio, heEPJ06}. In the left panel of Fig.~\ref{fig-4} we show the spin-averaged $1p$ and $1d$ $\bar{p}$ binding energies and widths for better comparison with the results obtained with the central Paris potential.\\

In conclusion, we performed self-consistent calculations of $\bar{p}$-nuclear quasi-bound states using a microscopic potential, namely the Paris $\bar{N}N$ potential, for the first time. We explored the effect of the $P$-wave interaction on $\bar{p}$ binding energies and widths. We found that the $P$-wave interaction almost does not affect the binding energies of ${\bar p}$-nuclear states. This is in sharp contrast to the case of ${\bar p}$ atoms where it was found necessary to include the $P$-wave interaction in order to increase attraction of the ${\bar p}$ optical potential~\cite{friedmanNPA15}. Moreover, we found good agreement between the results obtained using the phenomenological RMF potential and the Paris $S$-wave + phenomenological $P$-wave potential which are the two potentials consistent with antiprotonic atom data and $\bar{p}$ scattering off nuclei at low energies.

\section*{Acknowledgements}
We thank E. Friedman and A. Gal for valuable discussions, and B. Loiseau for providing us with the ${\bar N}N$ amplitudes.
This work was supported by the GACR Grant No. P203/15/04301S.

\end{document}